\documentclass{ws-rv9x6}
\usepackage{subfigure}   
\usepackage{ws-rv-thm}   

\usepackage[square]{ws-rv-van}
\makeindex
\def\bi{\bibitem}
\def\be{\begin{eqnarray}}
\def\ee{\end{eqnarray}}

\def\la{\langle}\def\ra{\rangle}\def\del{\partial}

\begin{document}
\chapter[ In Search of a Pristine Signal for (Scale-)Chiral Symmetry in Nuclei]{In Search of a Pristine Signal  \\  for (Scale-)Chiral Symmetry in Nuclei}\label{ra_ch1}
\author[M. Rho]{Mannque Rho}

\address{ Institut de Physique Th\'eorique, CEA Saclay, 91191 Gif-sur-Yvette, France \\
mannque.rho@cea.fr}

\begin{abstract}
\noindent  I describe the long-standing search for a ``smoking-gun" signal for the manifestation of (scale-)chiral symmetry in nuclear interactions. It  is prompted by Gerry Brown's last unpublished note, reproduced verbatim below, on the preeminent role of pions and vector ($\rho$,$\omega$) mesons in providing a simple and elegant description of strongly correlated nuclear interactions.  In this note written in tribute to Gerry Brown, I first describe a case of an unambiguous signal in axial-charge transitions in nuclei and then combine his ideas with the more recent development on the role of hidden symmetries in nuclear physics. What transpires is the surprising conclusion that the Landau-Migdal fixed point interaction $G_0^\prime$, the nuclear tensor forces and Brown-Rho scaling,  all encoded in scale-invariant hidden local symmetry,  as Gerry put, ``run the show and make all forces equal."
\end{abstract}

\body


\section{Introduction}
The currently active theoretical nuclear physics research is to calculate, ``ab initio," nuclear properties  in an effective field theory starting from chiral symmetry of QCD associated with the light-mass quarks relevant in nuclear interactions. This approach consists of calculating $m$-nucleon potentials for $m\geq 2$ with ``irreducible diagrams" to high orders $\kappa \gg 1$ in N$^\kappa$LO in  the standard chiral counting and computing  many-body nuclear correlations summing  ``reducible diagrams" in a variety of sophisticated many-body techniques. Here the term ``ab initio" refers then to the putative contact with QCD via effective field theory in the spirit of Weinberg's ``folk theorem"~\cite{weinberg-folk-theorem}.  In consistency with the folk theorem, the higher one can go up in $\kappa$, while preserving the required conditions such as symmetries etc., the better the calculation will fare in confronting Nature. Of course, given the nature of effective field theory, one is currently limited in scope by rapidly increasing number of parameters as $\kappa$ is increased, but unless the effective field theory in question breaks down -- which could happen under certain extreme conditions -- with more refined experimental information and increasing computer power, it is reasonable to expect that our understanding of what goes on in nuclear systems will be greatly improved in the years to come. One could say this is a nuclear physics proof of the ``folk theorem."

In this note, I would like to describe what Gerry and I undertook, initially in 1970's, then more intensively in 1980's and 1990's, to  uncover what we considered as ``preeminent features" of chiral symmetry, combined with a presumed scale symmetry, in QCD.  Looked at from the present state of art in high-order chiral effective field theory, most, if not all, of those features could very well be captured in high $\kappa$ calculations. Thus one might say, ``what's the big deal?"

Our philosophy has been this: Whatever beautiful phenomena there may be in the processes accessed -- and however well they are described -- by high-order and consistent chiral perturbation theory, they are likely buried and difficult to single out in full-fledged high-$\kappa$ computations. By identifying the preeminent features by relying on simplicity and intuition, one can reveal in what elegant way Nature works and make certain predictions that are hidden in  potentially accurate ``ab initio" approaches.

\section{Low-Energy Theorems}
What one might identify as the first ``signal" for chiral symmetry was ``seen" in the 1972 calculation by Riska and Brown for the thermal $np$ capture $n+p\rightarrow d+\gamma$~\cite{riska-brown} where the dominant source for  $\sim 10 \%$ meson-exchange effects in the cross section was identified.  This calculation was prompted by the observation that the soft-pion theorems given by the current algebra relations, fairly well established then, could give an important contribution to the M1 matrix element figuring in low-energy processes of the $np$-capture type~\cite{chemtob-rho}. Soon afterwards, it was realized that the soft-pion theorems must play an even more important role in weak processes in nuclei. Indeed it was predicted that the exchange of a nearly zero-mass pion could give a lot bigger contribution to the exchange axial-charge matrix elements in nuclear beta decay than in the M1 process~\cite{kubodera-delorme-rho}. This prediction was confirmed, convincingly, by several experiments of first-forbidden beta decay transitions as I will describe below (see for an early discussion, Ref.~\cite{BR-comments}).   The corollary to these two (confirmed) observations was that both the EM charge operators and the weak Gamow-Teller operators could not  receive contributions from soft-pion exchanges and hence must be subject to higher-order corrections. These observations were made before Weinberg's 1979 paper~\cite{weinberg-chpt} established that these current algebra terms are the first term in the chiral perturbation expansion, which is the source of the subsequent developments in EFT for QCD and the current lively activity in the nuclear physics community.  In the modern language, therefore, the soft-pion terms operative in the M1 and axial-charge operators are the leading exchange-current operators with  next-order corrections suppressed  by one or two (in the latter case) chiral orders.   This was already understood in 1981 when we started~\cite{BR-comments,MR-Erice} to formulate the chiral counting rule in nuclear interactions motivated by Weinberg's 1979 paper. It is perhaps fair to say that our work foresaw the arrival of nuclear EFT largely triggered by Weinberg's influential paper on nuclear chiral effective field theory~\cite{weinberg-nuclearEFT}.\footnote{The formulation that we initiated in early 1980's was interrupted by the rediscovery in the context of QCD of the Skyrme soliton model for nucleons in 1983 which took us away from our activity on chiral EFT, and was made more complete after Weinberg's paper. See \cite{MR-chpt}.}

As I shall stress below, the enhanced soft-pion exchange currents offer a {\it clear signal} of how chiral symmetry (more precisely scale-chiral symmetry, specified below) manifests in nuclei. This point, reinforced later in the context of what's known as ``BR scaling" as mentioned below, was underlined in our 1981 note~\cite{BR-comments} where it was stated ``Meson exchange currents, therefore, probe the structure of the `vacuum' inside the nucleus." I will come back to this matter to argue for a ``pristine" signal for scale-chiral symmetry.

\section{Scale-Chiral Symmetry}
\subsection{Vector mesons and scalar meson}
When Gerry Brown received the first draft of Weinberg's 1991 article with a request for comments, his first reaction was that he preferred that the vector-meson degrees of freedom, in particular, that of the $\rho$ meson, be explicit in the Lagrangian, instead of being generated at higher chiral orders as in the chiral Lagrangian with pions only, adopted by Weinberg and in almost all of the current applications. The reason for this was Gerry's conviction that certain properties of nuclear forces could be most economically and efficiently captured if the vector mesons were treated explicitly. He was persuaded on this by a variety of nuclear observables connected to the nuclear tensor force~\cite{BR-tensorforces}, which is one of the most important component of nuclear forces,  in particular,  spin-isospin response functions, the vector dominance and most significantly the effect of vacuum change in dense nuclear medium.

Although Gerry relied mostly on intuitive reasoning at the early stage, the most rigorous way presently available to address the problems involved is now recognized to be to resort to flavor gauge symmetry for the vector mesons $V=(\rho,\omega)$ supplemented by scale symmetry for a scalar degree of freedom, with the vector and scalar degrees of freedom making up the crucial ingredients of the argument.

It is now pretty convincingly established (as is summarized for example in the review \cite{HRW}) that up to nuclear matter density, most, if not all, of nuclear properties are well described by chiral EFT\footnote{By chiral EFT, I mean EFT based on chiral Lagrangian with pions only (with or without baryon fields).}.  This means that relevant fluctuations with the quantum numbers of vector and scalar mesons, if important, could be properly captured in higher-loop terms in chiral EFT.  However such an EFT must break down when the energy scale probed becomes  comparable to the mass of heavier mesons, possibly at some density above normal. This could happen if the ``effective" vector meson mass in medium went down as suggested in the structure of the tensor forces~\cite{BR-tensorforces}, and more seriously at high density if the mass went to near zero according to the  ``vector manifestation" (VM)~\cite{HY:PR}.  One way the dropping vector meson mass can be handled is to treat  the vector meson as a local gauge boson.  In fact, it is only in gauge symmetry that one can take the $\rho$ meson as light and gets, at the leading order, the KSRF relations and vector dominance that agree well with experiments. In addition, one can set up a chiral perturbative scheme, if the vector meson mass is formally considered as light as the pion mass (although it is  $\sim 6$ times the pion mass in the vacuum), with a systematic chiral expansion~\cite{HY:PR} that works fairly well in the vacuum. In medium it would work even better, the lighter the $\rho$ mass dropped as predicted in HLS at high density. There are only two cases known in gauge theory where the notion of ``light" (strong-interaction) vector meson makes sense; one is the case of hidden gauge symmetry we are dealing with and the other is a supersymmetric QCD in some special parameter space~\cite{komargodski}.

An equally important degree of freedom in nuclear physics is a scalar meson of mass $\sim 600$ MeV that {\it effectively} provides the  attraction that binds nuclei. In the particle data booklet, there is a scalar of comparable mass, namely, $f_0(500)$, with a large width. In the view Gerry and I have advocated since 1991, the scalar is a dilaton resulting from spontaneously broken scale symmetry. We consider it as ``light" in the same sense as the $\rho$ mass ($m_\rho\approx 770$ MeV) is ``light." This is an assumption that of course needs still to be confirmed by higher-order calculations (in the scheme mentioned below).  In fact it is a long-standing controversy, with no clear consensus,  whether such a scalar -- that we will denote as $\phi$ -- can be associated with scale invariance.  On the one hand, lattice calculations indicate there can be an infrared (IR) fixed point within the conformal window but at large number of flavors $N_f\sim 8$. This is the case for which the Higgs may be identified as a dilaton~\cite{yamawaki-higgs}. An active work on this issue is in progress in anticipation of further discoveries at LHCb~\cite{yamawaki-higgs,yamawaki-technirho}. But so far there is no firm nonperturbative evidence, lattice or otherwise,  for an infrared (IR) fixed point for $N_f\leq 3$ that we are concerned with in QCD. There is therefore a school -- call it ``no-go school" -- that dismisses the notion of a dilaton scalar for $f_0(500)$. On the other hand, there is a conjecture that $f_0(500)$ could be interpreted as a dilaton with an IR fixed point with the $\beta$ function for the QCD gauge coupling $\alpha_s(=g_{QCD}^2/4\pi)$ vanishing and hence the trace of the energy-momentum tensor $\theta_\mu^\mu$ vanishing in the chiral limit~\cite{CT}. In this scheme, scale symmetry and chiral symmetry merge into what is called ``scale-chiral symmetry"~\cite{CT} with their scales locked to each other, $4\pi f_\chi\approx 4\pi f_\pi$ where $f_\chi$ is the dilaton decay constant. At present, there is no rigorous no-go theorem against scale-chiral symmetry either, so that possibility cannot be ruled out on theoretical ground. Furthermore this scheme has a great advantage not only for particle physics (such as, among others, giving a simple explanation for the famous $\Delta I=1/2$ rule\footnote{This is somewhat like the status of the KSRF relations before the notion of hidden local symmetry was introduced.}) but also for nuclear physics where the dilaton scalar $\phi$  can provide a systematic scale-chiral expansion including a scalar meson, generalizing the standard chiral expansion. It can provide justification to the long-standing use -- with success -- of a local scalar field  for nuclear potentials (e.g., Bonn potential),  Walecka-type mean-field models etc. In addition, it offers an additional procedure to calculate its mass, width etc. at low loop-orders both in and out of medium. Perhaps more importantly, it would provide a more efficient method to do calculations where strange hadrons, such as hyperons and  kaons, relevant for compact-star matter -- as suggested in the counting rule  in \cite{CT} -- are involved.   We will see below that this scalar as a dilaton plays a key role in what Gerry and I have been doing all along.

The question remains, however, as to whether the failure for the lattice calculations  to ``see" the putative IR fixed point will not invalidate what I will be discussing below.  I have no clear answer to this.  In my opinion, one way to address this issue is to view the scale symmetry we are exploiting is an emergent symmetry in a way analogous to hidden local symmetry in baryonic medium. The $U_A (1)$ anomaly offers another analogy.

The ``no-go school" argument against a possible IR fixed point in QCD is anchored on the trace anomaly which cannot be turned off in the vacuum. The trace anomaly is due to the regularization required for the quantum theory, or put differently, the dimensional transmutation, and is renormalization-group invariant\footnote{I am grateful to Koichi Yamawaki for his point of view on this matter.}. Similarly the $U_A(1)$ anomaly cannot be turned off. It can be tuned to zero if the number of colors $N_c$ is tuned to $\infty$. However, in Nature,  $N_c\ll \infty$, so the axial anomaly is there to stay. Nonetheless it has been argued that the $U_A(1)$ symmetry could be restored at high temperature~\cite{pisarski-wilczek}.\footnote{Up to date lattice calculations fail to see the phase transition up to $\sim 1.5$ times the chiral transition temperature for $N_f=2$~\cite{karsch}.}  In a similar vein, it is possible that the trace anomaly could be turned off, in the chiral limit, by density with the symmetry exposed at some high density. This possibility is being explored~\cite{PKLR}. Since I am not concerned here with densities much higher than that of normal nuclear matter, I will not go into it.
\subsection{Hidden symmetries}
It should be recognized that {\it both} the local symmetry for the $\rho$  and the scale symmetry for the $\phi$ are ``hidden" symmetries: They are not visible or may even be absent in QCD proper.   The hidden local symmetry for the $\rho$ becomes manifest only when the $\rho$ mass is driven toward -- but not exactly onto -- zero~\cite{HY:PR}\footnote{Unless otherwise noted, I will be working with the chiral limit.}.  Since this flavor local symmetry is not present in QCD proper, the $m_\rho=0$ limit may not be accessible in QCD.  I will however suggest that it can emerge via strong nuclear correlations in dense medium.  As for scale symmetry, it can be shown that the familiar linear sigma model  has the scale symmetry {\it hidden} in it.   It has been shown~\cite{yamawaki-technirho} that by dialing one parameter $\lambda$ in a potential term in the standard linear sigma model (which is equivalent to the standard Higgs model) from $\infty$ to 0, the Lagrangian can go from the non-linear sigma model with no conformal (scale) symmetry to a conformal-invariant nonlinear sigma model. With hidden local gauge fields suitably incorporated, the latter turns into scale-invariant hidden local symmetry (sHLS for short). Below I will use sHLS with the scale symmetry spontaneously broken by a potential $V(\chi)$ where $\chi$ is what is referred to as ``conformal compensator field" connected to the dilaton $\phi$ (defined below).  What the baryon density does is to drive the parameter between  $\lambda=\infty$ and $\lambda=0$ and expose the hidden symmetries somewhere along the way.
\section{Scale-Invariant Hidden Local Symmetric Nuclear EFT}
Let me start with the mesonic Lagrangian denoted as sHLS that combines scale symmetry and hidden local symmetry that can be written in a schematic form:
\be
{\cal L}_{sHLS}={\cal L}_0 (U, \chi, V_\mu) +{\cal L}_{SB} (U,\chi)\label{sHLS}
\ee
where the conformal compensator field $\chi$ is related to the dilaton field $\phi$ as
\be
\chi=f_\chi e^{\phi/f_\chi}
\ee
and the chiral field is given by the familiar form $U=e^{i\frac{2\pi}{f_\pi}}$.  In this article I will deal with flavor $SU_f(2)$ and assume $U(2)$ symmetry for the vector mesons $V=(\rho,\omega)$.\footnote{There is a strong indication that this symmetry is badly broken in sHLS at a density denoted $n_{1/2}\sim (2-3)n_0$ where $n_0$ stands for normal nuclear matter density. I will not deal with this high density which is relevant for compact stars~\cite{PKLR}.}  The mass dimension-one $\chi$ transforms linearly under scale transformation while $\phi$ transforms nonlinearly as the pion field $\pi$ does under chiral transformation, that is, as a Nambu-Goldstone. The first term (\ref{sHLS}) is of scale dimension 4, so gives scale-invariant action and also HLS (chiral) invariant. The second term contains the pseudo-scalar meson mass term, hence breaking explicitly the chiral symmetry, and a potential that breaks scale symmetry both explicitly (due to the trace anomaly) and spontaneously. Although most of my discussions in application to nuclei can be done in the chiral limit, the chiral symmetry breaking pion term will be necessary to fix the property of the pion decay constant which sets the density-scaling behavior in nuclear matter.

Expanded in fields, the Lagrangian (\ref{sHLS}) in unitary gauge in hidden gauge symmetry  is, to ${\cal O}(p^2)$ chiral order, of the form
\be
{\cal L}_{\chi_s} &=& \frac{f_\pi^2}{4}\kappa^2
\mbox{Tr}\left(\partial_\mu U^\dagger \partial^\mu U\right) + \kappa^3 v^3 \mbox{Tr}{\cal M}\left(U+U^\dagger\right)
\nonumber\\
& &{} - \frac{f_\pi^2}{4} a\kappa^2 \mbox{Tr}\left[\ell_\mu + r_\mu + i(g/2)
( \vec{\tau}\cdot\vec{\rho}_\mu + \omega_\mu)\right]^2 \nonumber\\
& &{} - \textstyle \frac{1}{4} \displaystyle
\vec{\rho}_{\mu\nu} \cdot \vec{\rho}^{\mu\nu}
-\textstyle \frac{1}{4}  \omega_{\mu\nu} \omega^{\mu\nu}
+\frac 12 \del_\mu\chi\del^\mu\chi + V(\chi)
\label{shls}
\ee
where $a$ is related to the ratio $f_\pi/f_\sigma$ (where $f_\pi$ is the pion decay constant and $f_\sigma$ is the decay constant of the would-be Goldstone boson Higgsed to give the vector meson mass), $\kappa=\chi/f_{\chi}$ with $f_{\chi}=\la 0|\chi|0\ra$, $l_\mu=\del_\mu\xi\xi^\dagger$, $r_\mu=\del_\mu\xi^\dagger\xi$ with $\xi=\sqrt{U}$ and $v$ is a constant of mass dimension 1.
\subsection{Intrinsic density dependence}
In order to apply (\ref{sHLS}) to baryonic matter, baryon degrees of freedom are needed. There are two ways to bring in baryon fields. One way, perhaps most consistent with QCD, is to generate baryons as skyrmions of the mesonic Lagrangian. There is a work along this line with some progress. At present,  however,  it is not developed well enough to quantitatively describe nuclear processes. The alternative is to put baryon fields explicitly in consistency with the folk theorem, staying as faithful as possible to the symmetries involved. I will follow this approach  below. Let me call the baryon-implemented effective Lagrangian bsHLS.

Since the ultimate aim is to probe the density regime where the vector mesons, i.e., $\rho$ and $\omega$, and the scalar are relevant, perhaps overlapping the regime where the explicit QCD degrees of freedom may intervene, the strategy to take is to have the EFT matched to QCD at the scale where the cutoff for the EFT is set. In \cite{HY:PR}, the matching was done by means of the vector and axial-vector correlators, tree order in HLS (i.e., ``bare" HLS) and OPE in QCD, at $\Lambda_M=\Lambda_\chi\sim 4\pi f_\pi$.  With sHLS, the energy-momentum tensor needs also to be matched. What the matching does is then to endow the ``bare" parameters of the EFT Lagrangian with dependence on the condensates, i.e., the quark condensate $\la\bar{q}q\ra$, the gluon condensate $\la G^2\ra$ and mixed forms etc. Since those condensates depend on the vacuum, if the vacuum is modified by density, then they will necessarily depend on density. This dependence will then render the parameters of the Lagrangian {\it intrinsically} density-dependent. This density dependence, of QCD origin, is called ``intrinsic density dependence" (IDD for short).

Given the Lagrangian so matched to QCD, then one has an effective Lagrangian, the ``bare" parameters of which are  density-dependent, with which one can do quantum theory. The IDD so defined is related -- but not identical -- to what is known in the literature as ``Brown-Rho scaling" (BR scaling for short)\footnote{The BR scaling as applied to certain nuclear processes may contain other density dependence than IDD. For example, when short-range contact 3-body forces are integrated out, the resulting two-body force can inherit the three-body force effect in the form of BR scaling. One can understand this by that the zero-range three-body forces figuring in chiral EFT, involving $\omega$ and heavier meson exchanges of bsHLS,  are of the same or higher scale than the cutoff scale $\tilde{\Lambda}$, hence their effects get captured in BR scaling when integrated out to arrive at chiral EFT. This will be the case with the C14 dating problem mentioned below. This means that the BR scaling used there has a contribution from the three-body force effect in addition to IDD. Another example: The axial current coupling $g_A$ is different for Gamow-Teller transitions (space component of the current) from axial charge transitions, also discussed below. The former is BR and the latter is IDD.  Note that IDD in EFT Lagrangian is a Lorentz-invariant object while BR in physical observables may contain Lorentz-breaking contributions. A most prominent example where BR scaling in practice can contain more than  IDD is the anomalous orbital gyromagnetic ratio $\delta g_l$ which can be well described by a BR scaling expressed in terms of certain Fermi-liquid parameters. It is in Fermi-liquid parameters that the IDD is lodged~\cite{friman-rho}. Here the link between BR scaling and IDD is indirect and complicated.}.

The most efficient and flexible approach, presently available, to treat many-body nuclear dynamics with bsHLS is the renormalization-group approach employing $V_{lowk}$.  It involves ``double decimations"~\cite{BR:dd}.  For nuclear processes, one should be able to do the decimation from a cutoff $\tilde{\Lambda}$ somewhat lower than $\rho$ mass. From the vector meson mass scale, the ``bare" parameters of the EFT Lagrangian -- except for $f_\pi$\footnote{The pion decay constant does, however, flow by pion loops below $\tilde{\Lambda}$.}\label{6} -- do not flow to the scale picked, $\tilde{\Lambda}$, from which the decimation is to be done, so it should be justified to lower the cutoff to $\tilde{\Lambda}$ without modifying the ``bare" Lagrangian. In practice, the first decimation is made from $\sim (2-3)$ fm$^{-1}$ to obtain the  $V_{lowk}$ and then the 2nd decimation consists of doing Fermi-liquid calculations with this $V_{lowk}$~\cite{holt-fermiliquid} .

In HLS taken to ${\cal{O}}(p^2)$ in the chiral (derivative) counting, there are only three parameters $g$, $f_\pi$ and $a$. In the skyrmion description, nucleon properties including couplings to the vector mesons involved do not require additional parameters. With the scalar field included, there is of course an additional parameter, namely,   $f_\chi$. However the locking of scale symmetry and chiral symmetry makes $f_\chi$ equal to $f_\pi$, so it does not require additional IDD.

Now the question is how these parameters vary as a function of density and how their dependence affects hadron masses and coupling constants of the ``bare" Lagrangian?

The answer to this question requires knowing whether there is any phase change in the matter structure as density increases. It is obvious that the parameters will not necessarily continue moving smoothly in density. For example, at some density, QCD degrees of freedom could enter. In the skyrmion description of baryonic matter, there is a robust topological transition from a skyrmion matter to a half-skyrmion matter at a density around $n_{1/2} \sim (2-3)n_0$. In fact this transition is in a sense equivalent to what is called ``quarkyonic" in which quark degrees of freedom figure at about the same density~\cite{fukushima,PKLR}.  In terms of the ``bare" Lagrangian, such a transition would imply changes in the density dependence of the bare parameters. I won't go into what happens after such transition which matters for compact star structure -- which has been studied~\cite{PKLR}, so let me focus on the density regime $n<n_{1/2}$.

First consider the $\rho$ mass. The ``bare" mass at the matching scale $\Lambda_M\sim \Lambda_\chi$ is given by
\be
m_\rho^2=af_\pi^2 g^2.\label{ksrf}
\ee
What is remarkable about this relation, known as KSRF formula, is that it holds to all orders of loop corrections with the HLS Lagrangian taken to ${\cal O}(p^2)$ -- and believed to be valid at higher chiral (derivative) orders -- with corrections of ${\cal O}(m_\rho^2/\Lambda_\chi^2)$~\cite{ksrf-allorders,HY:PR} .  This expression therefore becomes more accurate, the lighter the vector mass becomes as is predicted in dense medium~\cite{HY:PR}. This means that the ``bare" $\rho$ mass in the Lagrangian will always be of the form of (\ref{ksrf}), regardless of the cutoff for decimation, with the IDD reflecting {\it entirely} the effect of density. The explicit calculation of  the EFT-QCD matching formulas shows that both $g$ and $a$ depend quite weakly on the quark and gluon condensates~\cite{HY:PR} and hence the  density dependence will be mainly in the pion decay constant, hence in the dilaton condensate since $f_\pi\approx f_\chi$.  Therefore the only scaling factor in the density regime $n\leq n_{1/2}$ is
\be
f_\pi^\ast/f_\pi\approx f_\chi^\ast/f_\chi\equiv \Phi (n).
\ee
Thus via (\ref{ksrf})
\be
m_\rho^\ast/m_\rho \approx \Phi.
\ee
It follows from the bare Lagrangian (\ref{shls}) with the expansion $\chi=\la 0^\ast|\chi|0^\ast\ra +\chi^\prime$ (where $0^*$ is the in-medium vacuum)  that
\be
m_H^\ast/m_H\approx \Phi(n)\label{BR}
\ee
for $H=N,\rho,\omega,\phi$ assuming $U(2)$ symmetry for $(\rho,\omega)$ for $n<n_{1/2}$. The pion mass, with  broken chiral symmetry -- and hence broken scale symmetry, scales differently,
\be
m_\pi^\ast/m_\pi\approx \sqrt{\Phi}.\label{BRpi}
\ee
Equations (\ref{BR}) and (\ref{BRpi}) are the same as the expressions derived in 1991 using the skyrmion model. Note that they follow not directly from chiral symmetry but from scale symmetry locked to chiral symmetry.  In other words, {\it it is the dilaton condensate that ``runs" the show}. This means that the symmetry involved is the ``scale-chiral symmetry" as defined precisely in \cite{CT}.

\subsection{Soft-pion signal for scale-chiral symmetry}
The first ``pristine" signal for scale-chiral symmetry in nuclei is in the axial-charge beta decay process in nuclei. It is in the first-forbidden beta transition of the form
\be
I(0^-)\rightarrow F(0^+) +e^- +\bar{\nu}_e\ \ \Delta T=1.
\ee
This beta decay process from the initial nucleus $I$ to the final nucleus $F$ goes via the axial change operator $A_0^a$.

\begin{figure}[ht]
\begin{center}
{\includegraphics[width=6cm]{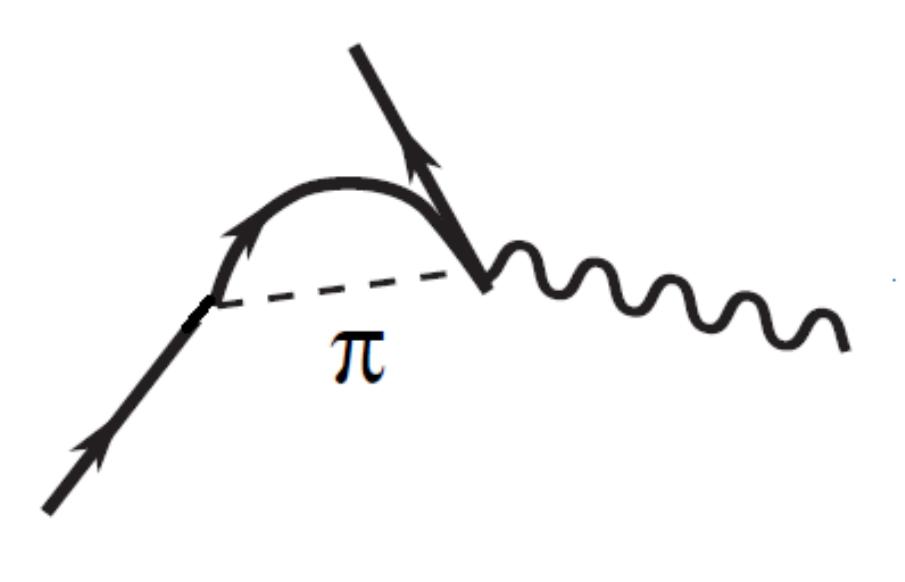}}
\vskip -0.5cm
\caption{Effective single-particle soft-pion-exchange axial charge operator. The solid line is the nucleon and the wiggly line the  external weak field. The right vertex $A_\mu^a \pi$NN is large for the time component $A_0^a$ and is suppressed for the space component $\mathbf{A}^a$.}\label{axial-charge}
\end{center}
\end{figure}

As first recognized in 1978~\cite{kubodera-delorme-rho} from current algebras and later confirmed in chiral perturbation theory~\cite{MR-chpt}, the exchange axial-charge two-body operator receives a large contribution from a soft-pion exchange term, with the next contribution suppressed by two chiral order. Furthermore the leading one-body operator, being first-forbidden, is kinematically suppressed. Therefore the two-body ``correction" term is expected to contribute to the decay at an order comparable to or bigger than the ``leading" single-particle operator. Written in effective one-body operator, the corresponding Feynman diagram is of the form Fig.~\ref{axial-charge}.
The sum of the one-body and two-body axial charge operators with the IDD incorporated into the constants of the EFT Lagrangian has the extremely simple form~\cite{kk-br}
\be
A_0^a=g_A\frac{\tau^a}{2}\frac{\mathbf{\sigma}\cdot\mathbf{p}}{m_N\Phi} (1+\frac{R}{\Phi})
\ee
where $\mathbf{p}$ is the nucleon momentum,  $R$ is the ratio of the matrix element of the two-body operator ($M_2$) to that of one-body operator ($M_1$). The factor $\Phi$ corresponds to the IDD dependence in the EFT Lagrangian. It may be that the numerical values of $M_1$ and $M_2$ depend on how nuclear wave functions are calculated. However the ratio $R$ is highly insensitive to it. Thus one can take either Fermi-liquid model or Fermi-gas model in place of more sophisticated wave functions. One gets essentially the same value. What is significant is that the $R$ is big $R\sim {\cal O}(1)$, and varies slowly in density, reflecting the robust nature of the soft-pion exchange.

One can readily make a simple estimate of what comes out. Let's look at the quantity defined and measured experimentally by Warburton~\cite{warburton}
\be
\epsilon_{MEC}=\frac{1}{\Phi} (1+\frac{R}{\Phi}).\label{epsilon}
\ee
This represents the enhancement factor due to both the exchange-current contribution in the transition matrix element relative to the single-particle operator contribution  {\it and} the IDD.
Take the lead $A=205-212$  nuclei for which data are available which have densities comparable to nuclear matter density. Calculating $R$ in Fermi-gas model, it is found at nuclear matter density $n_0$ that $R(n_0) \approx 0.5$. Now from pionic atom data, one has $\Phi (n_0)\approx 0.8$~\cite{yamazaki}. Thus $(\ref{epsilon})$  gives $\epsilon_{MEC}\approx 2.0$. This agrees very well with the measured enhancement factor  $\epsilon^{exp}_{MEC}=2.01\pm 0.05$. Taking into account the density dependence of $\Phi$ and $R$, one can also reproduce the observed enhancements in $A=12$ and $A=16$ systems, $\epsilon^{exp}_{MEC} (A=12)=1.64\pm 0.05$~\cite{minamisono} and $\epsilon^{exp}_{MEC} (A=16)\approx 1.7$~\cite{garvey}.  Although the theoretical estimate as well as the experimental values for $\epsilon_{MEC}$ are rough, this is a clear evidence for both the soft-pion and IDD effects: The density dependence of $\epsilon_{MEC}$ is found to be consistent with what's predicted of $\Phi$ and $R$.

Two remarks are in order here.

One is the crucial role of soft pions. It zeroes in on the Nambu-Golddstone-boson nature of the pion with a mass nearly vanishing (on the strong interaction scale). For very low-energy processes, $E\ll m_\pi\approx 140$ MeV, according to the standard lore of EFT, one may be justified to integrate out the pions leaving only the nucleons as relevant degrees of freedom. One then gets  what is called ``pionless effective field theory" ($\not{\pi}$EFT), which is generally thought to be  consistent with the ``folk theorem." With the pions absent, the resulting Lagrangian is blind to  chiral symmetry but it does not mean chiral symmetry is violated. So does it always work? The question is: Is $\not{\pi}$EFT applicable to the axial-charge transition which receives big contributions from soft pions?   If the pion mass were strictly zero for which the soft-pion theorems hold, clearly the pion could not be integrated out from the chiral Lagrangian. Thus it seems inevitable that the $\not{\pi}$EFT Lagrangian with the  pions gotten rid of would miss the soft-pion effect (at the $\kappa=1$ order) and hence fail even if it were applicable to Gamow-Teller transitions.

The other remark is on the possibility to do precision calculations and test the combined enhancement by soft pions and IDD. The operators are well defined to the leading chiral order with higher-order terms strongly suppressed, so given accurate wave functions, one could then do a precision test of the scaling parameter $\Phi$. Recent developments on ``ab initio" approaches with sophisticated many-body techniques could be exploited to calculate $\epsilon_{MEC}$ with accurate error estimates for both theory and experiment.
\section{What runs the show in nuclear interactions?}
Let me now come to the startling, if not puzzling, observation -- the main thrust of Gerry's note -- that the $\pi$, $\rho$, $\omega$ and $\phi$, the principal degrees of freedom of bsHLS at mean-field, play the dominant and even clear-cut role in nuclear dynamics. For this part,  reading Gerry's note (added below) will be helpful.

 Given the bsHLS Lagrangian,  one could perhaps perform a (covariant) density functional analysis  for the nuclear ground-state properties along the line set up by the Hohenberg-Kohn theorem for atomic/molecular physics and chemical physics.   The currently popular covariant energy-density functional approaches employed in nuclear theory typically have six or more free parameters. In contrast, up to at least nuclear matter density, the bsHLS Lagrangian was found to have basically only one parameter \footnote{Or at most three if fine-tuning is needed  to obtain a precision fit to data. Since the only parameter of the theory, $\Phi$, can be fixed by experiments at least up to nuclear matter density,  there is no real free parameter in the theory.}  governing the d-scaling factor $\Phi$ associated with the dilaton condensate. It would be extremely interesting to see how an ``ab initio" covariant density functional given by  bsHLS compares with the standard approach with many more parameters.  However one is ultimately interested in the equation of state relevant to compact stars. For this the density functional approach does not seem appropriate.

Instead the strategy followed below is the double-decimation renormalization-group (RG) procedure in terms of $V_{lowk}$~\cite{BR:dd}, which is to start with the first RG decimation to go from the effective cutoff $\tilde{\Lambda}$ down to the scale at which the $V_{lowk}$ is gotten. For this, in principle, the ``irreducible graphs" are to be summed to high orders in scale-chiral expansion to give the potential with which to do the decimation. In doing this, the IDDs enter. In practice, BonnS-type potentials are used and the cutoff is put at (2-3) fm$^{-1}$, somewhat lower than $\tilde{\Lambda}$. Performing the second decimation corresponds to doing Landau Fermi-liquid theory with $V_{lowk}$ as formulated in \cite{vlowk,holt-fermiliquid}. The result is then the set of Fermi-liquid fixed point parameters, i.e., the effective masses and interactions. I will focus on the Fermi-liquid parameters and the fixed point quasiparticle interactions in which  the pion and the $\rho$ play the main role, in particular, the Landau parameter $G_0^\prime$ and the tensor forces.
\subsection{The EELL effect or $G_0^\prime$}
In his note, Gerry Brown presents strong intuitive arguments, drawing from the previous works~\cite{vlowk,holt-fermiliquid}, to show that the Landau parameter $G_0^\prime$, coming from the pion and $\rho$ exchanges, is by far the largest among the Fermi-liquid interactions and dominates the Kuo-Brown effective interactions at mean-field level, with higher order terms suppressed~\cite{holtetal}. How and to what extent the suppression takes place more generally is yet to be worked out. However in the Wilsonian RG approach, the beta function for the quasiparticle interactions should tend to zero in the large $N$ limit where $N$ is related to the Fermi momentum $k_F$, as beautifully explained in \cite{shankar}. Gerry relies on the double-decimation $V_{lowk}$ approach with BR scaling implemented to argue that the Kuo-Brown interaction, with just one bubble, which is classical in nature, has the most of the physics in it, giving an extremely simple interpretation of why and how the high-order core polarization contributions are suppressed as found in \cite{holtetal}. Both the scalar $\phi$ and the vector $\omega$ figure in bringing the interaction $\sim G_0^\prime (\tau_1\cdot\tau_2)(\sigma_1\cdot\sigma_2)$ to the fixed point, the former ``holding the ball of pions together" and the latter providing the short-range repulsion giving rise to the Ericson-Ericson-Lorentz-Lorenz effect.
\subsection{Nuclear tensor forces}
The exchange of $\pi$ and $\rho$ with the parameters endowed with IDD of the EFT Lagrangian gives the effective in-medium nuclear tensor forces. While the pion tensor is more or less unaffected by density, an effect which could be attributed to protection by chiral symmetry, the $\rho$ tensor, with the dropping mass, increases in magnitude with a sign opposite to that of the pion tensor as density increases. Because of the cancellation between the two, the net tensor force strength gets weaker due to the BR scaling at increasing density. The attraction in the tensor channel goes nearly to zero when density reaches 2$n_0$.  This has been well known. In fact precisely this effect has been exploited with an impressive success to explain the long C14 life-time~\cite{holt-c14}. What's involved in this process is a delicate density-dependent cancelation in the Gamow-Teller matrix element, which nearly vanishes in the density regime involved. This is a spectacular signal of working of the BR scaling, although it is not as clear-cut an evidence as in the case of the first-forbidden beta decay process described above. As noted by Gerry, in the context of double decimation, the effect of BR scaling here is equivalent to the effect of contact three-body forces in the sense described in footnote 8.

I will now propose that there is a possibility of ``seeing" the IDD scaling -- not just BR scaling -- in nuclei via the tensor forces. I would like to describe this using the $V_{lowk}$ formalism.

In a series of beautiful papers~\cite{otsuka}, Takaharu Otsuka showed that the tensor forces played a remarkable role in the ``monopole" matrix element of the two-body interaction between two single-particle states labeled $j$ and $j^\prime$ and total two-particle isospin $T$
\be
V_{j,j^\prime}^T=
\frac{\sum_J (2J+1)\la jj^\prime|V|jj^\prime\ra_{JT}}{\sum_J(2J+1)}.
\ee
What is special with this matrix element is that it affects the evolution of single-particle energy;
\be
\Delta\epsilon_p (j)=\frac 12(V_{jj^\prime}^{T=0} +V_{jj^\prime}^{T=1})n_n(j^\prime)
\ee
where  $\Delta\epsilon_p (j)$ represents the change of the single-particle energy of protons in the state $j$ when  $n_n(j^\prime)$ neutrons occupy the state $j^\prime$. It turns out that the matrix elements $V_{jj^\prime}$ and $V_{j^\prime j}$ have opposite sign for the tensor forces if $j$ and $j^\prime$ are spin-orbit partners.

Let me summarize the salient features of the shell evolution connected to the tensor force found by Otsuka:

Otsuka works with the phenomenological potential Av18' and the one given by ChPT at N$^3$LO, both treated \`a la  $V_{\rm lowk}$. Other realistic potentials are found to give the same results. He varies the cutoff $\tilde{\Lambda}$ and finds $\tilde{\Lambda}$ independence around 2.1 ${\rm fm}^{-1}$.

\begin{figure}[htb]
\begin{center}
  \includegraphics[width=8cm,clip]{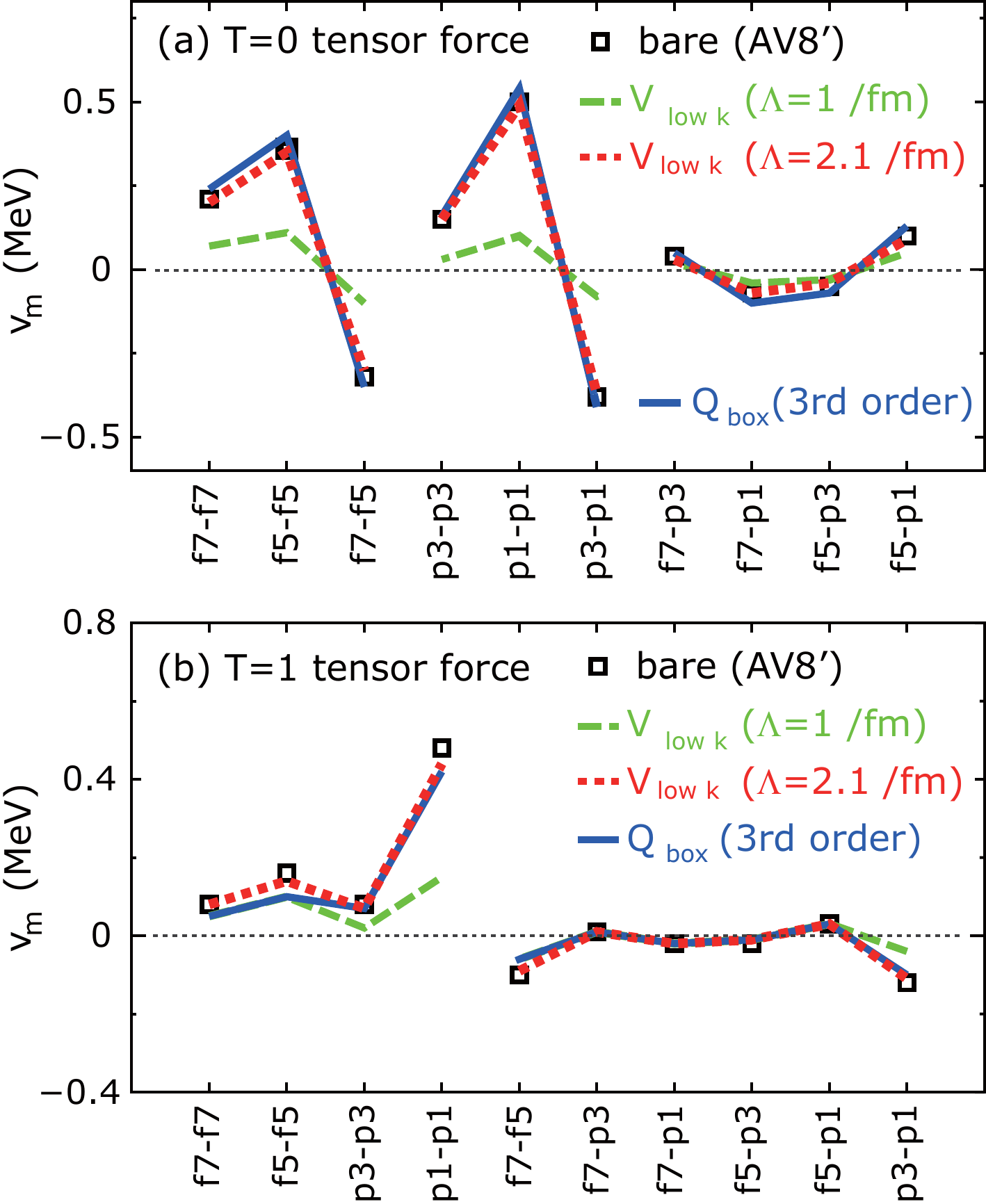}
 \caption{Tensor forces in AV8' interaction, in low-momentum interactions in the pf shell
obtained from AV8', and in the 3rd-order Q$_{box}$ interaction for (a) T=0 and
(b) T=1.
}
  \label{pfshell}
 \end{center}
\end{figure}

 Otsuka calculated the shell evolution in the pf and sd regions by including high-order correlations using the  Q-box formalism to 3rd order. While he finds the central part of the potential strongly renormalized by high-order terms, the tensor forces are left unrenormalized, leaving the ``bare" tensors more or less intact.
Shown  in Fig.~\ref{pfshell} is Otsuka's result (copied from his paper) in the pf shell region.
The result shows that the sum of the short-range correlation and medium effects as taken into account by the 3rd order Q-box leaves the bare tensor force unchanged. Otsuka looks at a variety of other realistic potentials, both phenomenological as well as ChPT at N$^3$LO, and finds, remarkably, that they all  give the same result, showing that the effect is robust.  It implies
\be
\frac{d}{d\tilde{\Lambda}} V_{low k}^{tensor}=\beta([V_{low k}^{tensor}], \tilde{\Lambda}) \approx 0.\label{beta}
\ee
Some experimental data are available, e.g., Jahn-Teller effect~\cite{otsuka},  that verify the tensor implemented calculations to be in agreement with experimental data quite well. Forthcoming experiments in RIB accelerators promise to reveal more surprising results.

\begin{figure}[h]
 \begin{center}
 { \includegraphics[width=5.5cm,clip,angle=0]{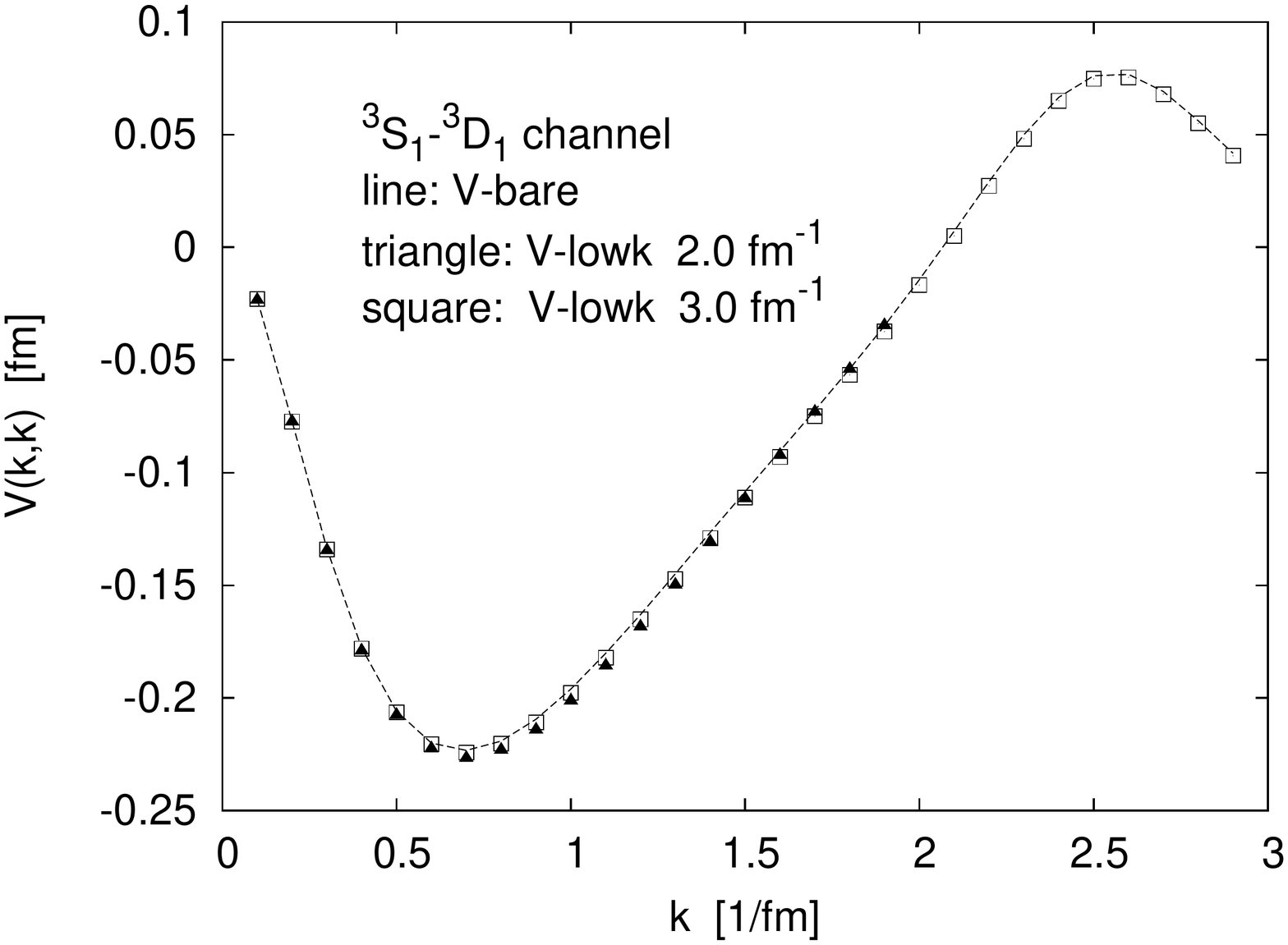} \includegraphics[width=5.5cm,clip,angle=0]{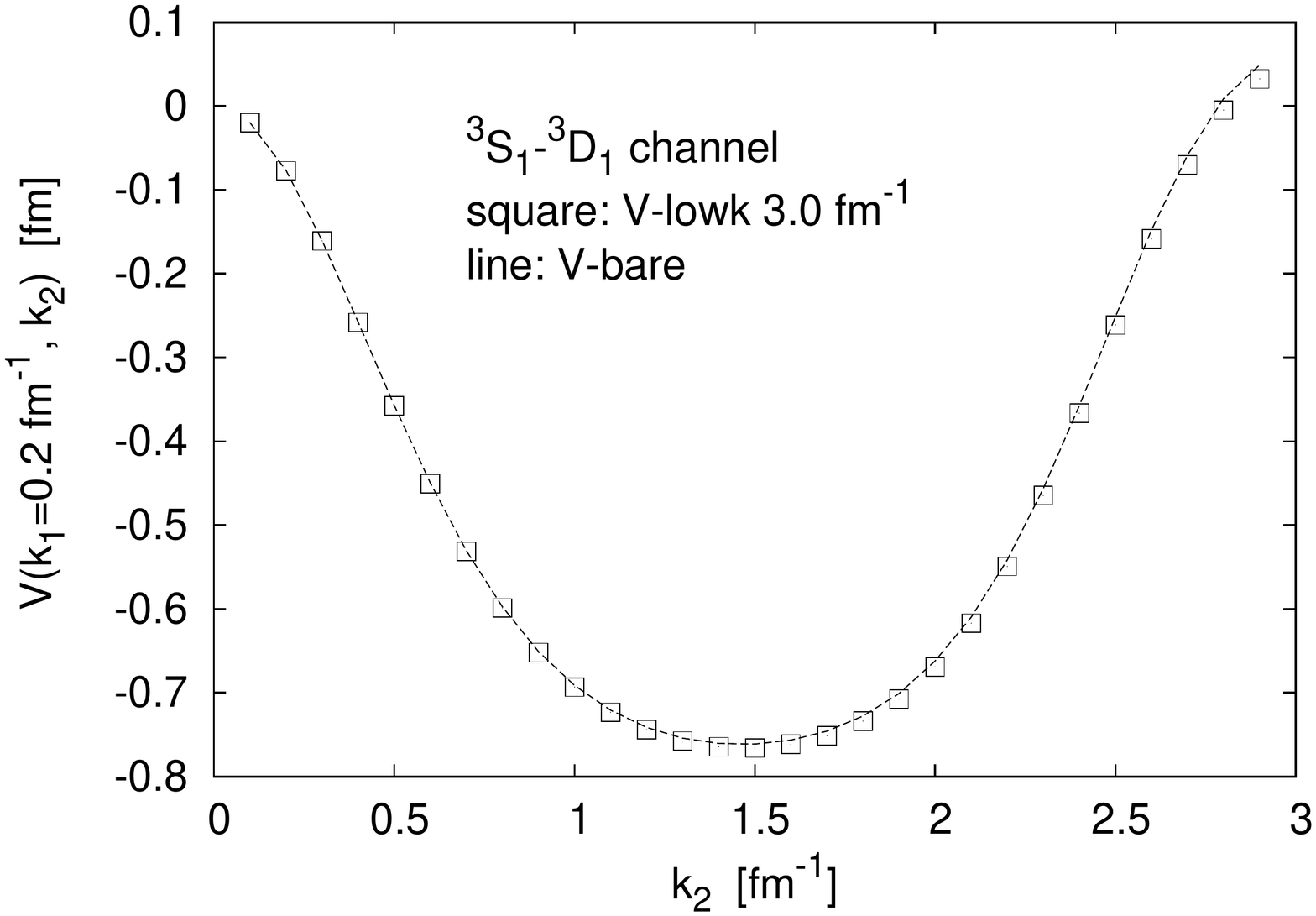}}
 \caption{``Bare" BonnS and $V_{low k}$ tensor forces  in matter-free space. Courtesy of Tom Kuo. }
  \label{tensor-free}
 \end{center}
\end{figure}

The result (\ref{beta}) says that the beta function is zero {\it both} in the first decimation and in the second. The latter could perhaps be understood as the net tensor force at a given density being at the Landau fixed point with all correlation effects suppressed. However the former is surprising since it implies that the tensor force does not RG-flow in the vacuum. Why the net tensor force is free of all strong interaction effects, in and out of medium, is mysterious. Learning of Otsuka's results, Tom Kuo kindly performed a $V_{lowk}$ analysis in the vacuum.  Shown in Fig.~\ref{tensor-free} are his results of the tensor potentials in momentum space $V^{tensor} (k_1,k_2)$ for $k_1=k_2$ and $k_1\neq k_2$.   The $V^{tensor}_{lowk}$ potential is identical to the ``bare" BonnS tensor potential, independent of $\tilde{\Lambda}$ .

\section{Seeing ``pristine" signals}
Let me close this note with what ``pristine" signal could mean for chiral symmetry (or more precisely scale-chiral symmetry).  At the early stage of dilepton production in relativistic heavy-ion collisions, Gerry imagined that it would mean the $\rho$ mass going to zero at the critical temperature $T_c$ (in the chiral limit). Gerry's idea was that in medium the $\rho$ mass, rather than the quark condensate, is the relevant order parameter to measure for chiral restoration, and in HLS which is a natural framework for addressing the issue, this meant going toward the VM fixed point as $T_c$ is approached.  However the VM fixed point with an enhanced symmetry turns out not to be in QCD in the vacuum~\cite{georgi,HY:PR}. It thus could make sense -- if at all -- only as an emergent phenomenon. Furthermore HLS predicts that in approaching the VM fixed point, while the $\rho$ mass could go to zero with a vanishing width, the photon tends to largely decouple from the dropping-mass $\rho$. This is quite unlike seeing the $\rho$ in the vacuum by measuring the dileptons of the given invariant mass of $\rho$ with a detector in the vacuum. In our view, dileptons in heavy-ion processes are not at all a suitable probe for scale-chiral symmetry: the $\rho$ meson produced in the process  is so strongly distorted by background nuclear correlations in the medium so that the signal for the $\rho$ meson carrying the order parameter subject to the VM, when measured with a detector outside of the medium, will be like a ``needle in the haystack"  as Gerry and I,  with our collaborators,  have argued.

So how does one go about ``seeing" the signal?

I have argued in this note that ``seeing" (scale-)chiral symmetry in action in nuclei is much like ``seeing" the meson-exchange currents in nuclei. Three such signals for scale-chiral symmetry in nuclei are described.

The first is the combined effect of soft pions and IDD, giving a whopping factor of $\sim 4$ effect in the decay rate. It reveals {\it both} the presence of meson-exchange currents {\it and} the influence of scale-chiral symmetry manifested by the excitation of pions.  It further shows that pions cannot be integrated out for certain processes -- such as first-forbidden transitions - that single out soft-pion dominated effects in near zero-energy processes.

The second is the Landau parameter $G_0^\prime$ dominated by the $\pi$ and $\rho$-meson exchanges in strong correlations with the scalar and $\omega$ mesons  (binding and short-range) which in Gerry's words, runs the ``show" in nuclear dynamics.

The third is the non-renormalization of the net tensor force in and out of medium, offering the possibility of a pristine probing for IDD by zeroing in on those processes that are controlled by the (net) tensor force and by dialing the density of the system. Of course how to dial the density is a big open issue.

Finally all these are simple and elegant aspects of nuclear interactions which could be sharpened in ``ab initio" precision calculations -- in progress and to come.
\subsection*{Acknowledgments}
I am grateful for fruitful discussions and collaborations with Masa Harada, Tom Kuo, Hyun Kyu Lee, Yong-Liang Ma and Won-Gi Paeng and would like to acknowledge extensive comments and tutorials from Rod Crewther, Lewis Tunstall and Koichi Yamawaki on scale symmetry in gauge theory. I would particularly like to thank Tom Kuo for his help on the nuclear tensor force at its putative fixed point.

\newpage
\setcounter{equation}{0}

\centerline{\bf \large Note Added: Unpublished Paper by G.E. Brown }

\centerline{\bf \large  ``The Dominant Role of the EELL Interaction in Nuclear Structure" }

\begin{figure}[ht]
\begin{center}
\includegraphics[width=6.7cm,angle=-1.5]{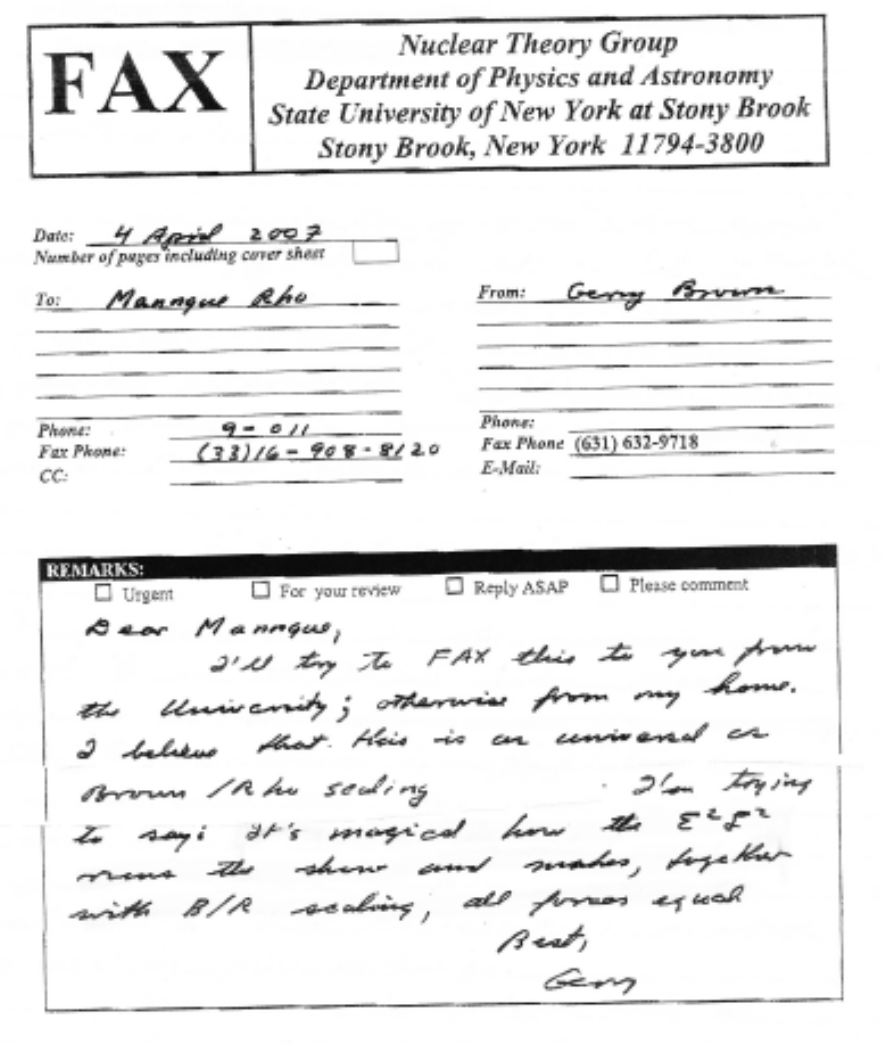}\includegraphics[width=6.5cm]{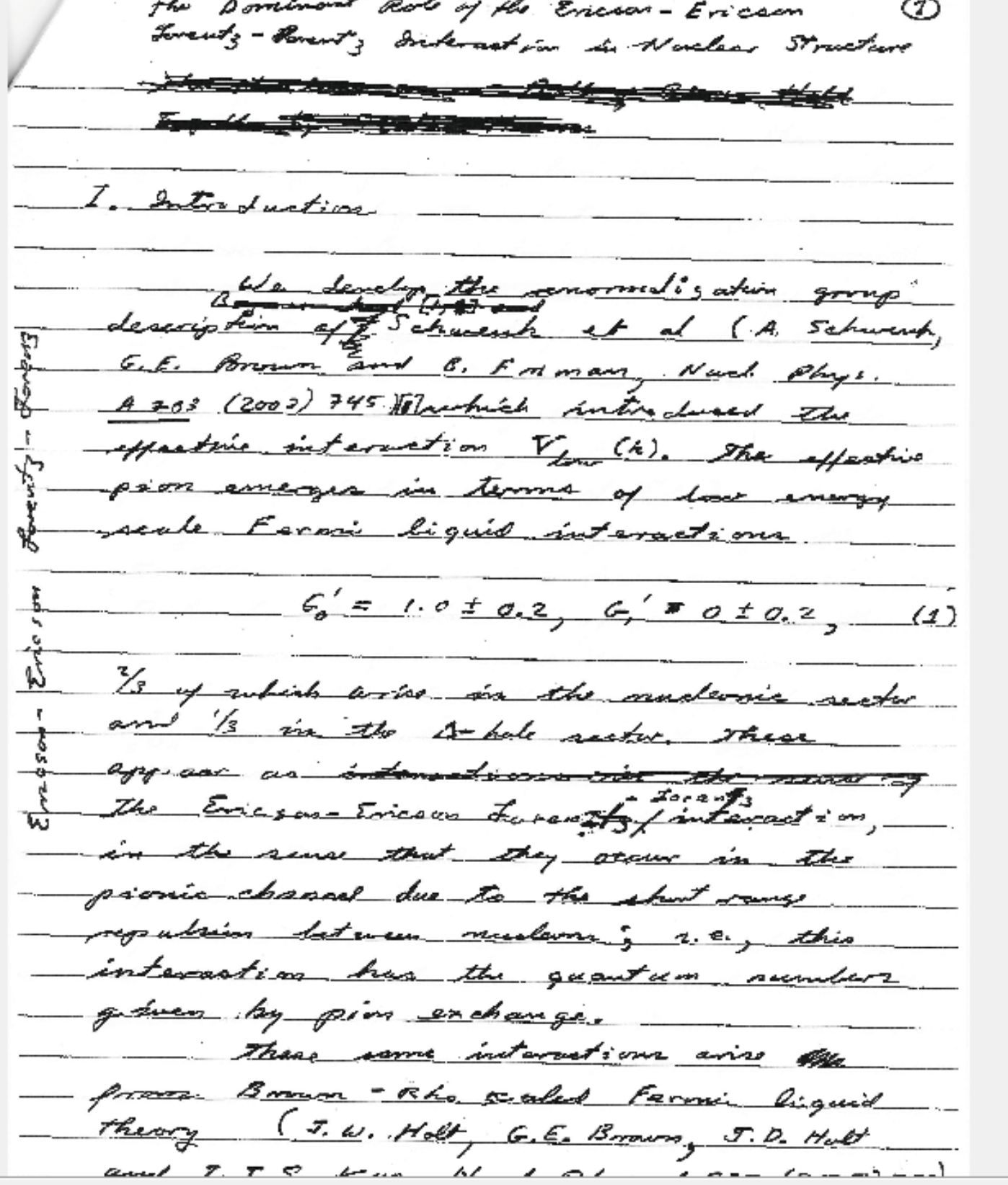}
\caption{Cover letter (left) and title page (right) of the faxed paper.}  \label{fax}
\end{center}
\end{figure}

\vskip 1cm
\centerline{\bf Forewords}
\vskip 0.5cm
{\it This is Gerry's last manuscript, hand-written and as he used to do, faxed to me on April 4th, 2007, the cover letter and the first page of which are scanned and put in Fig. \ref{fax}. Below, the article is transcribed verbatim, totally unedited and unrevised. It contained no abstract, but a short statement on the cover letter conveyed the essence of the idea.  Referring to the Ericson-Ericson-Lorentz-Lorenz (EELL or E$^2$L$^2$ for short) interaction, which captures the RG fixed point interaction $g_0^\prime$ in Landau-Migdal Fermi liquid theory, Gerry stated ``I believe this is as universal as Brown-Rho scaling. It's magical how E$^2$L$^2$ runs the show and makes, together with B-R scaling, all forces equal." In the contribution to this volume given above, I  describe how Gerry's ideas can be synthesized into the more recently developed notion of scale-invariant hidden local symmetry in nuclear interactions.}

\subsection*{Introduction}
We develop the renormalization group description of Schwenk et al.~\cite{schwenk} which introduced the effective interaction $V_{lowK}$, The effective pion energies in terms of low energy scale Fermi-liquid interactions
\be
G_0^\prime=1.0\pm 0.2, \ \ G_1^\prime=0\pm 0.2\,
\ee
2/3 of which arise in the nucleonic sector and 1/3 in the $\Delta$-hole sector. These appear as the Ericson-Ericson Lorentiz-Lorenz interaction, in the sense that they occur in the pionic channel due to the short-range repulsion between nucleons; i.e., this inteaction has the quantum numbers given by pion exchange.

These same interactions arise from Brown-Rho scaled Fermi-liquid theory~\cite{holt} as repulsive interactions between nucleons, replacing the attractive piece of the zero range pion-nucleon interaction which is removed by the short-range repulsion between nucleons. The effective $\vec{\sigma}_1\cdot\vec{\sigma}_2 \mathbf{\tau}_1\cdot\mathbf{\tau}_2$ component of the interaction found by Holt et al.~\cite{holt} is not only the strongest component of the effective interaction, but is exactly that found by Schwenk et al.~\cite{schwenk}. The long-range scalar exchange through the $\sigma$-meson and coupled two-pion exchange tends to hold the ball of pions together.

As far as we can see, were we to include the $\Delta$ with coupling constant $g_{N\Delta}^\prime=0.3$ and the double $\Delta$ with $g_{\Delta\Delta}^\prime=0.3$ then we would have a description in the double-decimation decimation of Brown and Rho~\cite{BR:DD} dual to that of Schwenk, Brown and Friman~\cite{schwenk}  of the screening by the $\Delta$-hole channel of the screening of the interaction in the pionic interaction. We return to this later.

\subsection*{Nuclear Matter with Brown-Rho Scaled Fermi-liquid Interactons}
We believe that it is useful to review the work of Holt et al.~\cite{holt} quantitatively in comparison with that of Schwenk et al.~\cite{schwenk}  because the double decimation is essentially the same as the latter with introduction of three-body forces. In any case, the substantial improvement in going over to the double decimation from the usual Fermi-liquid approach in the long wave-length limit should serve a stimulus for pursuing this direction. We reproduce Table 4 of Ref.\cite{holt} as Tabel 1 below.
\begin{table}[ht]
\tbl{Nuclear observables from the self-consistent solution obtained by iterating the Babu-Brown equations.}
{\begin{tabular}{@{}ccccc@{}}
\hline
  & $V_{NI}$ & $V_{NII}$ &$V_{N93}$  & $V_{CDB}$ \\
\hline
$m^\star/m$ & 0.721  & 0.763 & 0.696 &0.682 \\
$K$ MeV &218 &142&190 &495 \\
$\beta$ MeV & 20.4  & 25.5 & 23.7 &19.2 \\
$\delta g_l$ & 0.296  & 0.181 & 0.283 &0.267 \\
\hline
\end{tabular}}\label{table1}

\end{table}

In Table \ref{table1} we show the effective mass, compression modulus, symmetry energy, and anomalous orbital gyromagnetic ratio for the Nijmegen I ($V_{NI}$) and II ($V_{NII}$) and CD-Bonn ($V_{CDB}$) potentials with the in-medium modification \`a la Brown-Rho. We also show for comparison the results from the Nijmegen 93 ($V_{N93}$) one-boson exchange potential, which has only 15 parameters and is not fine-tuned separately in each partial wave. The iterative solution is in better agreement with all nuclear variables.  The anomalously large compression modulus in the CD-Bonn potential results almost completely from the presence of an $\omega$ coupling $g_{\omega NN}^2/4\pi=20$ as discussed in \cite{holt}. Otherwise we do not see much difference between the generally good fit to observables.

Since the double decimation, although crudely done, generally gives observables close to the empirical ones we shall not try to distinguish between them and following Ref.\cite{holt} we take the average of them and quote a deviation, reproducing Table 5 of \cite{holt} as Table 2 below.
\begin{table}[ht]
\tbl{Fermi liquid coefficients for the self-consistent solution to the Babu-Brown equations. }
 %
{\begin{tabular}{ccccc} 
\hline
 $l$ & $F_l$ & $G_l$ &$F_l^\prime$  & $G_l^\prime$ \\
\hline
0 & $-0.20\pm 0.39$  & $0.04\pm 0.11$ & $0.24\pm 0.16$ & $0.53\pm 0.09$ \\
1 & $-0.86\pm 0.10$ & $0.19\pm 0.06$ & $0.18\pm 0.05$ & $0.17\pm 0.01$ \\
2 & $-0.21\pm 0.01$  & $0.12\pm 0.01$ & $0.10\pm 0.02$ & $0.01\pm 0.02$ \\
3 & $ -0.09\pm 0.01$  & $0.05\pm 0.01$ & $0.05\pm 0.01$ & $0.01\pm 0.01$ \\
\hline
\end{tabular}}\label{table2}
\end{table}

Aside from the very large compression modulus $K=495$ MeV in the CDB the variuous observables are not significantly different from each other.

The results of Holt et al. differ from those of Schwenk et al. in that they do not include the $\Delta$ isobar. The latter renormalize $G_0^\prime$  to take effects from the $\Delta$ into account.. Without their effects, Kawahigashi et al.~\cite{kawa}  find $g_{NN}^\prime=0.6$ to compare with $G_0^\prime=0.53\pm 0.09$ from our Table 2. Generalization by Schwenk et al. to a model corrected for the screening due to $\Delta$-hole excitations to all orders with $NN\rightarrow N\Delta$  and $N\Delta\rightarrow N\Delta$ interaction strengths of $g_{N\Delta}^\prime=0.3$ and $g_{\Delta\Delta}^\prime=0.3$ by K\"orfgen et al.~\cite{kor1} gives $G_0^\prime=1.0$. Thus, we believe that adding the $\Delta$ to our description will have the same consequene and that we are starting with a $g_{NN}^\prime$ close enough so that we have essentially the same screening by a small change in our $\Delta$ coupling.
\subsection*{$G_0^\prime$ as the Main Interaction Strength}
Were there no two-body correlation function keeping the two nucleons apart, there would be no pionic interaction with them in the long wavelength limit since it is derivative in nature
\be
\delta H=\frac{f}{4\pi} \bar{\psi} \mathbf{\sigma}\cdot \mathbf{\nabla}\vec{\tau}\cdot\vec{\pi}\psi (r)
\ee
and the momentum $\vec{p}$
\be
\mathbf{\nabla}\vec{\pi}=\mathbf{p}\vec{\pi}
\ee
goes to zero as the volume goes to $\infty$. The total interaction by way of pion exchange is
\be
V(r)= \frac{f^2}{r}\frac 13 \mathbf{\sigma}_1\cdot\mathbf{\sigma}_2 \vec{\tau}_1\cdot\vec{\tau}_2\Big(\frac{e^{-m_\pi r}}{m_\pi r}-\frac{4\pi}{m_\pi^3}\delta (r)\Big)\label{pipot}
\ee
plus a tensor interaction which averages to zero over angles. The integration of the above $V(r)$
\be
\int d^3r V(r)=0
\ee
does go to zero. The Ericsons~\cite{ericsons} included short-range correlations by multiplying $V(r)$ by a short-range correlation function $g(r_{12})$ which had the property that
\be
g(r_{12})=0, \ \ r_{12}=0
\ee
but otherwise was of sufficiently short range that it did not affect anything.

If we leave the $\delta(r)$ out we have a minimalist description of the effect of short-range correlations. We are not finished because it is well-known that the exchange of $\rho$-mesons with tensor coupling between two nucleons contributes to the Lorentz-Lorenz effect. Brown~\cite{brown} found that inclusion of this contribution increases the pionic one by a factor of 1.8, increasing the 1/3 in Eq.~\ref{pipot} to 0.6. We note that this is close to the $G_0^\prime$, the average $\sigma\tau$ Fermi liquid interaction in the $l=0$ state of the interactions shown in
Table 2. Furthemore, the 0.6 is precisely what Kawahigashi et al.~\cite{kawa} find for the contribution to the E$^2$L$^2$ interaction from the nucleon channel alone.

We see that the $G_0^\prime=1$, $G_1^\prime=0$ is the largest interaction by far. (It will be clear below why we group them.) In Table 2 the next largest ones are $F_l=-0.20$ and $F_l^\prime=0.24$. We divide $F_l$ by 3 because of its angle dependence.

We see that $G_0^\prime=1$, $G_1^\prime=0$ implies a $\delta$-function interaction,  the potential
\be
V(r)=\frac{\pi^2}{2k_f/m^\star} \mathbf{\sigma}_1\cdot \mathbf{\sigma}_2 \vec{\tau}_1\cdot\vec{\tau}_2 \delta(r)
\ee
representing the potential energy necessary to pull the nucleon and antinucleon apart from each other to overcome the pionic attraction. Note that this potential is always attractive, because the two fermions are at the same point and since they are antisymmetrical, if they are spin triplet they must be isospin singlet and vice versa.

Note that $V_{lowk}$ gives quite a good description of the $^{18}$O and $^{18}$F spectra~\cite{holt2}. These calculations do not have the Brown-Rho-scaled masses in them, however, and would be expected to change as have results of Table 1 and Table 2.
\subsection*{Conclusion}
We believe that we can offer a qualitative understanding of one of the main points of nuclear structure physics without detailed calculations; namely, why the one-bubble correction to the mean-field (shell-model) spectrum is very important as is well known in the Kuo-Brown interactions, whereas higher order corrections, as found in Holt et al.~\cite{holt2} do not change the pattern qualitatively. This is because the E$^2$L$^2$  interaction is by far the strongest one and it is largely spent in the single bubble. In preliminary estimates we find that the direct and exchange effects largely cancel each other in two-bubble corrections and suggest that the one bubble which is classical in nature has most of the physics in it. The last paragraph is highly speculative.

\end{document}